\documentclass[preprint,12pt,authoryear]{elsarticle}

\usepackage{algorithmic}
\usepackage{graphicx}
\usepackage{tcolorbox}
\usepackage{textcomp}
\usepackage{xcolor}
\usepackage{hyperref}

\journal{Journal of Systems and Software}

\begin{document}

\begin{frontmatter}



\title{A Study of Update Request Comments \\in Stack Overflow Answer Posts}


\author[inst1]{Mohammad Sadegh Sheikhaei}
\author[inst1]{Yuan Tian}

\affiliation[inst1]{organization={School of Computing, Queen's University},%Department and Organization
           % addressline={Goodwin Hall}, 
            city={Kingston},
            %postcode={K7L2N8}, 
            state={ON},
            country={Canada}}

\author[inst2]{Shaowei Wang}

\affiliation[inst2]{organization={Department of Computer Science, University of Manitoba},%Department and Organization
            %addressline={University of Manitoba}, 
            city={Winnipeg},
           % postcode={R3T2N2}, 
            state={MB},
            country={Canada}}

\begin{abstract}
Comments play an important role in updating Stack Overflow (SO) posts. They are used to point out a problem (e.g., obsolete answer and buggy code) in a SO answer or ask for more details about a proposed answer. We refer to this type of comment as update request comments (URCs), which may trigger an update to the answer post and thus improve its quality. 

In this study, we manually analyze a set of 384 sampled SO answer posts and their associated 1,221 comments to investigate the prevalence of URCs and how URCs are addressed. We find that around half of the analyzed comments are URCs. While 55.3\% of URCs are addressed within 24 hours, 36.5\% of URCs remain unaddressed after a year. Moreover, we find that the current community-vote mechanism could not differentiate URCs from non-URCs. Thus many URCs might not be aware by users who can address the issue or improve the answer quality. As a first step to enhance the awareness of URCs and support future research on URCs, we investigate the feasibility of URC detection by proposing a set of features extracted from different aspects of SO comments and using them to build supervised classifiers that can automatically identify URCs. Our experiments on 377 and 289 comments posted on answers to JavaScript and Python questions show that the proposed URC classifier can achieve an accuracy of 90\% and an AUC of 0.96, on average.
\end{abstract}


% \begin{highlights}
% \item Perform an empirical study on the update request comments in Stack Overflow
% \item Provide an annotated dataset of 1,221 comments posted on 384 answers to Java questions
% \item Propose a supervised-learning approach to detect URCs with an average accuracy of 90\% 
% %\item Our approach can be used in Stack Overflow to decrease the rate of unaddressed URCs
% \end{highlights}


\end{frontmatter}



%%       \citet{<label>} ==> Jones et al. (2015)
%%       \citep{<label>} ==> (Jones et al., 2015)

\section{Introduction}\label{sec:introduction}
Many developers utilize Stack Overflow (SO) to find solutions for their programming issues. With about 22 million questions and 33 million answers\footnote{https://stackexchange.com/sites}, SO is the largest Q\&A site for computer programming \citep{May-2019}. According to Jeff Atwood, one of the founders of SO, the goal of SO is not ``answer my question''  but ``let's collaboratively build an artifact that will benefit future coders''~\citep{Atwood-2019}. As a result, most of the answers (sometimes the questions) on SO are required to be continuously edited to maintain/improve their quality via resolving the textual/code issues (e.g., handle deprecated APIs and fix flawed code snippet) in the previous version~\citep{Zhang-2019,Wang-2018-2}.

Comments on the SO posts are the main channel for users to communicate and discuss the potential problems in the posts. SO encourages users to post comments in an answer post when they find an issue in the answer and ask for an update on the answer explicitly (e.g., ``Please replace method A with method B as A is deprecated.'') or implicitly (e.g., ``So when using \verb!ArrayList::new! the given key is inserted into the list?''). We refer to such comments as \textit{update request comments (URCs)} because they can potentially trigger an update in the corresponding answer and thus improve its quality.

The questions and issues mentioned in URCs may be addressed in the next comment(s) or body of the corresponding answer post, or even both of them. However, there is no guarantee for such URCs to be addressed. In SO, when a user writes a comment under an answer post, the system notifies the owner of the post, i.e., \textit{answer owner}, about the new comment. Then for each URC, the answer owner can address it either by updating the answer body or by writing a new comment to reply. However, if the answer owner does not handle the problem, the URC remains unaddressed until other users address it in a new comment or in the body of the answer (i.e., becoming an \textit{answer editor}). Prior studies find only a small portion of their collected comments resulted in an update in the corresponding answer post, and answer owners ignore many requests of answer update raised in comments~\citep{Soni-2019,Zhang-2019}. In other words, handling such URCs is still problematic, and relying solely on answer owners to maintain their answers maybe not be realistic. SO is a collaborative community and all users on SO are encouraged to maintain its answers via collaborative editing~\citep{Wang-2018-2}. Therefore, SO probably needs to attract eyeballs from the entire community to handle URCs. To alleviate the above problem, a first step is to have a deep understanding of URCs and investigate the possibility of developing an automated approach to identify such URCs so that they are visible to the community.

In this work, we first conduct an empirical study on URCs for having a deep understanding of URCs in terms of their prevalence, the percentages of URCs remained unaddressed, how URCs are addressed by the community members (i.e., addressed in answer post, addressed in the following comment(s), or addressed in both), how fast are URCs addressed, what is the contribution of different user roles in addressing URCs, which post part they choose to address URCs, and if comment votes can be used to distinguish URCs from non-URCs. 

To answer the above questions, we manually examined 1,221 comments from a statistical randomly sampled 384 answer posts of Java questions (questions with the $<$java$>$ tag). We observed that half of the analyzed comments are URCs. More interestingly, while 55.3\% of URCs are addressed within 24 hours, 36.5\% of URCs remain unaddressed after a year. We also found that the majority (80.1\%) of addressed URCs are addressed by the answer owner. Among addressed URCs, 88.7\% were addressed in the next comments, 33.3\% were addressed in the post body, and 22.1\% were addressed in both. By investigating the comment scores we realized that their scores are not good means to detect URCs. 

Our findings show that although URCs are prevalent and more than half addressed within 24 hours, many are ignored (remained unaddressed). Moreover, these URCs might not be visible to the community as they may not be highly voted. Therefore, we continue to explore the feasibility of automated URC detection, as the first step towards improving the awareness of URCs. We propose multi-dimension features, such as comment author role and comment relative time, that could potentially differentiate URCs from other comments based on our manual analysis of 1,221 comments. Then we employ these features in common supervised learning models to identify URCs. Specifically, we apply random forest~\citep{RF}, logistic regression~\citep{LR}, naive bayes~\citep{NB}, and also two deep learning models (a CNN model by \cite{Qu-2019} and a Transformer based model by \cite{gu-2021}), and train them by different inputs, i.e., the extracted features, TF-IDF/text, or extracted features + TF-IDF/text. We use the 1,221 annotated comments related to Java topic (\textit{Java comments}) to train these models, and then test them on comments extracted from JavaScirpt questions and Python questions to evaluate their performance and generalizability. 

Our experiments on 377 and 289 comments posted on answers of JavaScript and Python questions respectively show that the models that are based on the extracted features outperform TF-IDF and text based models with a large margin in terms of accuracy and AUC. Also, among the investigated models, the Transformer based model (using BERT) that gets text and the extracted features as its input, archives the highest performance, i.e., around 90\% accuracy and 0.96 AUC, indicating that URCs are highly predictable.

Our contributions include:
\begin{itemize}
\item Perform an empirical study on the update request comments in Stack Overflow.
\item Propose a supervised-learning approach that leverages our extracted features to detect URCs with the average accuracy of 90\%. This approach can be used in Stack Overflow to decrease the rate of unaddressed URCs.  
\item Provide an annotated dataset of 1,221 comments\footnote{available at https://doi.org/10.6084/m9.figshare.19382156} that are posted on randomly selected 384 answers to Java questions. We have provided three level of annotation: 1---If it is a URC, 2---For URCs, where is it addressed, i.e., in the next comments, in the post body, or both. 3---For URCs that are addressed by the following comments, which comment addresses it.
\end{itemize}

\section{Background}\label{sec:background}
As mentioned earlier, SO posts (either questions or answers) are continuously updated to address different issues such as resolving bugs, meeting time-related concerns (such as deprecated APIs), and simplification. According to Stack Overflow\footnote{https://stackoverflow.com/help/editing} ``\textit{Any user can propose edits, but not all edits are publicly visible immediately. If a user has less than 2,000 reputation, the suggested edit is placed in a review queue. Two accept or reject votes are required to remove the suggested edit from the queue and either apply the edit to the post or discard it. Users with more than 2,000 reputation are considered trusted community members and can edit posts without going through the review process.}" As a result, if there is a problem with a post, users usually write a comment on the post and ask for an update. Also, users may ask for clarification or question other related cases, resulting in the generalization of a solution.

\begin{figure*} [ht]
\centerline{\includegraphics[width=1\textwidth]{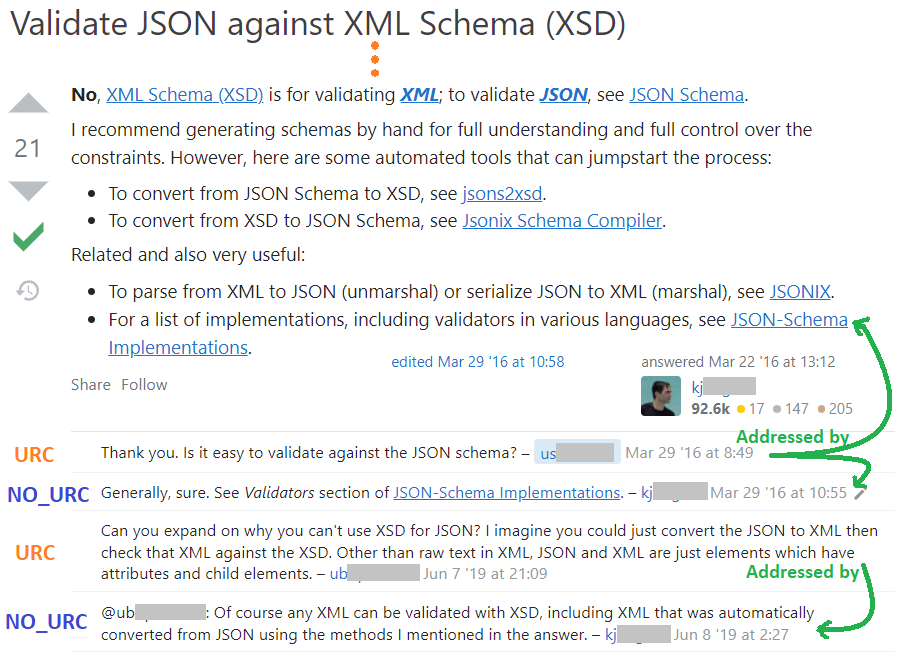}}
\caption{A sample answer post from a question tagged with ``Java''. There are two addressed update request comments and two non-update request comments on this answer.}
\label{fig-SOPostExample1}
\end{figure*}

\begin{figure*} [ht]
\centerline{\includegraphics[width=1\textwidth]{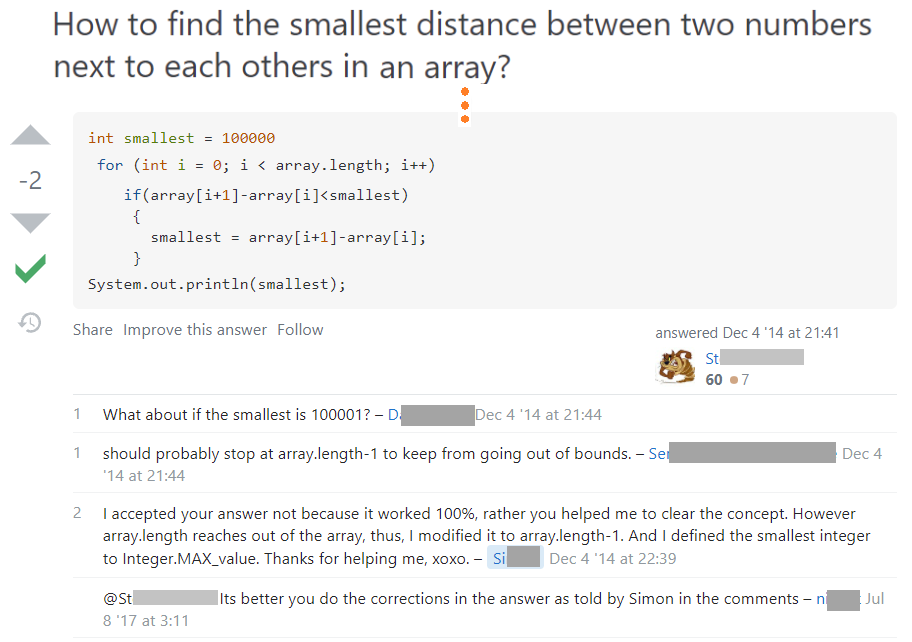}}
\caption{A sample answer posts from a question tagged with ``Java''. There are four unaddressed update request comments on this answer.}
\label{fig-SOPostExample2}
\end{figure*}

Fig.~\ref{fig-SOPostExample1} shows a sample answer post from a question having the $<$java$>$ tag along with its comments. There are four comments on this sample\footnote{https://stackoverflow.com/questions/36152972}. The first comment is written by the questioner (the user who started this page by posting the question) to ask more information about the answer. The second comment is written by the answerer (the user who wrote the answering post), to answer the first comment. The third comment is written by a third person to ask for more clarification, and the last comment is again from the answerer to address the third comment. Therefore, in this example, there are two URCs and two non-URCs. This sample also shows that there is an update on the answering post on Mar 29, 2016. By clicking on this date, SO opens a new page and shows the history of edits on this post by highlighting the new added parts in green and deleted parts in red. In this example, the last line of the answering post, i.e., ``\textit{For a list of implementations, including validators in various languages, see JSON-Schema Implementations.}", is appended by the answerer to address the first comment. Therefore, as shown in the figure, the first comment is a URC addressed in the next comment and the post body. But, the third comment is a URC that is only addressed by the next comment. Both of these URCs are addressed in less than 24 hours. Fig.~\ref{fig-SOPostExample2} shows another example\footnote{https://stackoverflow.com/questions/27304654} from the Java community. Although the answer is accepted, four unaddressed update request comments point out the problems with the proposed answer.

The effect of comments on answer updates was studied before by \cite{Soni-2019}.
%, but they didn’t provide an empirical study on different aspects of update request comments such as how fast they get addressed or \newtext{which user roles} addressed URCs and where they got addressed. In particular, \cite{Soni-2019} 
They employed three rule based heuristics to detect different types of comments regarding their effect on posts, although that approach discards around 35\% of comments due to not matching to some primary conditions. In contrast, we created a dataset of 1221 comments by providing three levels of annotation for each comment. We employed this dataset to (A) answer different empirical questions about URCs, (B) train various ML and DL models to predict the type of comments. Due to the differences between our approach and Soni’s work, we come to a significantly different conclusion on the ratio of addressed and unaddressed URCs. The details have been presented in Section~\ref{sec:relatedwork}.

\section{Empirical Analysis on Update Request Comments in SO Answers}\label{sec:empirical}
%\subsection{Research Questions and Study Design}

To understand the prevalence of update request comments (URCs) and how they are handled in practice, we first conduct an empirical study by answering the following four research questions. Answers to them would also guide future studies leveraging SO comments. As far as we know, this is the first empirical analysis of URCs.

\vspace{0.1cm}
\noindent \textbf{\textit{RQ1: How prevalent are URCs in technical Q\&A and how they get addressed by the community members?}} Knowing the percentage of URCs and the number of unaddressed URCs among all comments gives us insight into how critical it is to analyze URCs. Other information such as the ratio of URCs caused a post update gives us a higher level of perception about the user interactions via comments to improve the quality of answers.

\vspace{0.1cm}

\noindent \textbf{\textit{RQ2: How fast are URCs addressed?}} Knowing the URCs address speed gives us insight into the delay in addressing URCs, i.e., potential waiting time for SO users to get their URCs addressed. This metric could also help the SO community to keep track of the effectiveness of the community in addressing URCs. 

\vspace{0.1cm}
\noindent \textbf{\textit{RQ3: Which user role (questioner, answerer, other commenters) is more likely to address URCs? And in which part of the answer post do they choose to address URCs?}} Users who address URCs may have different roles in that SO page: questioner, answer owner, answer editor, and others. Also, they may address URCs in the post body, in the following comments, or both. Knowing which user role addressed URCs in which part of the answer post gives us insight on who contributes most to the address of URCs and where they prefer to address URCs.  

\vspace{0.1cm}

\noindent \textbf{\textit{RQ4: Can comment votes be used to distinguish URCs from non-URCs?}} SO uses a community-vote mechanism to decide which comments should be shown at top positions. Knowing if votes can help identify URCs gives us insight into whether URCs can be naturally selected by SO users or not.

\vspace{0.2cm}
To answer the above questions, we collect a set of comments and determine whether each of them is a URC or not. For each URC, we also denote if the URC is addressed or not. For addressed URC, we record if they are addressed in the answer post or in the following comments, and identify the role of SO user who addressed the URC. Section~\ref{sec:coding} and~\ref{sec:data} describe the coding guide for comment labeling and how we collect data for answering four RQs. In the end, we present the methodology and results of our empirical analysis in Section~\ref{sec:empiricalresult}.

\subsection{Coding Guide to Annotate the Comments}\label{sec:coding}

Before labeling the comments, we need a coding guide to annotate the comments by URC/NO\_URC, and then labeling the URC comments by URC\_ADDRESSED/ URC\_UNADDRESSED. A URC is a comment posted by any user but the answerer, explicitly or implicitly, asks to update the answer and improve its quality.  
As the answerer has the ability of changing his/her answering post, comments by this person is a NO\_URC comment. 
For the other comments that are from the questioner or third users, we check the content of the comment. If it either points out problems in the answer post, asks for more information that would help to understand the answer better, or provides important information (e.g., regulations) associated with the answer, we label it as URC (because it has the potential of initiating a post update), otherwise we label it as NO\_URC. Fig.~\ref{fig-DecisionTreeForManualLabeling} shows this process in a decision tree. According to this figure, the first and third comments in Fig.~\ref{fig-SOPostExample1} are URC because they ask for more information about a special case that would help the questioner to better understand the answer. However, the second and fourth comments are NO\_URC because they are written by the answerer. To provide more examples, these are some comments by a questioner or a third person on an answer post that we consider them as URC:

\begin{figure*} [ht]
\centerline{\includegraphics[width=0.8\textwidth]{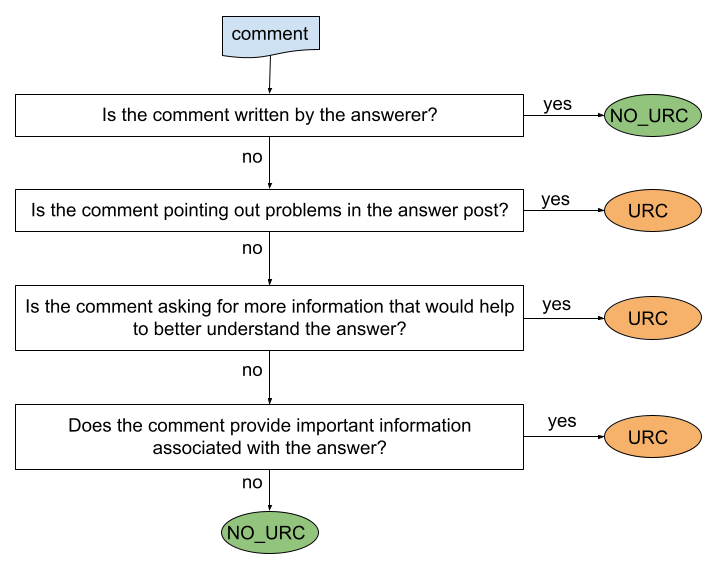}}
\caption{The decision tree applied in manually labeling comments by URC/NO\_URC}
\label{fig-DecisionTreeForManualLabeling}
\end{figure*}

\begin{itemize}
    \itemsep0em
    \item How can I use this solution in my code?
    \item I ran it and got this Exception: ...
    \item Can I use class B rather than class A in this solution?
    \item It doesn't work for me.
    \item Function A is deprecated.
    \item Works but runs slowly.
\end{itemize}

And here are examples of comments by a questioner or a third person on an answer post that we consider them as NO\_URC:

\begin{itemize}
    \itemsep0em
    \item Thank you!
    \item Great! This is the thing I was looking for.
    \item Oh, that's a nice point about having the possibility to put a null value there!
\end{itemize}

To tag the URCs with URC\_ADDRESSED/URC\_UNADDRESSED label, we add another level of annotation which determines if the URC is addressed in the following comments, or addressed in the post body, or addressed in both of them, or remained unaddressed. Any answering comment related to the URC (even if it is not the correct answer) is acceptable to tag that comment as URC\_ADDRESSED. We assume that if the answering comment is not what the user asked, she will write another URC. The only answering comments that we do not treat as answers are those that explicitly say ``I don't know'' and so forth.

To answer RQ2 and RQ3, we need to know which one of the next comments addresses the current URC (if any). Thus, the final step of our annotation process is to add another level of labeling to determine the addressing commentID (when the URC is addressed by the next comments). If there are multiple addressing comments, we take the first (the oldest) of them. As a result, our dataset has three levels (columns) of labeling, i.e., needs\_update, addressed\_in, and addressed\_by\_commentID. Table~\ref{table-labellingSampleComments} shows our labeling for the comments on the two answer posts mentioned in Fig.~\ref{fig-SOPostExample1} and \ref{fig-SOPostExample2}. Due to lack of space, only the first few words of each comment text is shown. In the ``user role'' column, we mentioned the role of the commenter. Refer to Table~\ref{table-comment-features} for more information about these roles. The three levels of our annotations are presented in ``needs update'', ``addressed in'', and ``addressed by commentID'' columns.

\begin{table*}[htbp]
\caption{Labels for the comments of two answer posts from SO Java community}
\vspace*{3mm}
\centering
\resizebox{14cm}{!}{%
\begin{tabular}{|c|c|c|c|c|c|c|c|}
\hline
\textbf{question} & \textbf{answer} & \textbf{comment} & & & \textbf{needs} & \textbf{addressed} & \textbf{addressed by} \\
\textbf{ID} & \textbf{ID} & \textbf{ID} & \textbf{user role} & \textbf{Comment text} & \textbf{update} & \textbf{in} & \textbf{comment ID}\\
\hline
\hline 
36152972 & 36155326 & 60185364 & Questioner & Thank you. Is it easy to ... & yes & both & 60190443 \\
\hline
36152972 & 36155326 & 60190443 & Answerer & Generally, sure. See ... & no & - & - \\
\hline
36152972 & 36155326 & 99591573 & Not seen commenter & Can you expand on why ... & yes & comment & 99594548 \\
\hline
36152972 & 36155326 & 99594548 & Answerer & @ubiquibacon: Of course any ... & no & - & - \\
\hline
\hline
27304556 & 27304654 & 43072230 & Not seen commenter & What about if the smallest ... & yes & no & - \\
\hline
27304556 & 27304654 & 43072237 & Not seen commenter & should probably stop at ... & yes & no & - \\
\hline
27304556 & 27304654 & 43073707 & Questioner & I accepted your answer not ... & yes & no & - \\
\hline
27304556 & 27304654 & 76939452 & Not seen commenter & @StevenAkaTaz Its better ... & yes & no & - \\
\hline
\end{tabular}
\label{table-labellingSampleComments}
}
\end{table*}

%%%%%%%%%%%%%%%%%%%%%%%%%%%%%%%

\subsection{Data Collection}\label{sec:data}
We collect data from the 2020\_03\_15 version of SOTorrent dataset \citep{Baltes-2019} by performing queries on Posts, PostHistory, Users, and Comments tables.

To achieve more reliable conclusions (less diverse results), in this study we focus on SO posts that are related to Java language (have \verb!<java>! tag). We expect to see similar results on other popular programming languages such as Python and JavaScript. We also consider only the questions that their score is equal or greater than zero. We assume that negative scored questions/answers have not attracted enough interest from the community to evaluate and classify their quality. As we aim to investigate the comments on answer posts, only the answers that have at least one comment are considered. Given the huge number of candidate answer posts, we focus on those that are recently touched, i.e., the last activity date (last post edit date, or post creation date if it is not edited) is 1/1/2017 or later. Moreover, the accepted answers or the answers that have the highest vote among answers posted on the same question are more important for the community. So, we only consider answers that either is accepted or has the highest vote. After filtering the answers based on the mentioned features, we ended up with 124,472 answers. Then, we randomly choose a sample of 384 answers to fulfill 5\% margin error with a confidence level of 95\% for our statistical analysis. To calculate the sample size, we apply the sample size calculator\footnote{ http://www.raosoft.com/samplesize.html}. Finally, we take all the comments of that 384 posts for our annotation process. Algorithm 1 shows the steps of data acquisition from SOTorrent.

\vspace{.2cm}

\noindent{
\begin{tabular}{p{0.95\linewidth}}
\hline
    \textbf{Algorithm 1} Data Acquisition from SOTorrent \\
\hline
    1- Select all questions with Java tag and $Score \ge 0$  \\
    2- Select answers of questions selected in Step 1 that A) is accepted or has the highest votes among answers posted on that question, B) posted or edited after 1/1/2017, and C) have one or more comments \\
    3- Randomly select 384 answers from the set of answers made in step 2 \\
    4- Select all comments of the 384 answers selected in the step 3  \\
\hline
\end{tabular}}

\vspace{.3cm}

To manually label the data, we follow the closed card sorting methodology. According to the coding guide for labeling the comments introduced in Section~\ref{sec:coding}, the first two authors manually examined 100 randomly sampled comments obtained from step 4, and only found six inconsistent annotations. They discussed and agreed on the final labels of the six comments. The first author then continues to label the rest comments. When annotating the comments, we found two answers with ambiguous comments. We ignored those two answers, and ended up with 1,221 annotated comments. Using that 100 randomly sampled comments that were annotated independently by the first two authors, the Cohen's kappa inter-rater agreement is 87.8\%, which indicates an excellent agreement. We will publish our dataset online for future research around this subject.

\subsection{Methodology and Results}\label{sec:empiricalresult}

\noindent \textbf{RQ1: How prevalent are URCs in technical Q\&A and how they get addressed by the community members?}

%Method
We use our labeled dataset to answer this question. Table~\ref{table-URC-statistics} shows the main statistical questions about update request comments. About half of the comments (631 of 1,221) are URC. Among 631 URCs, 417 comments (about 66\%) are addressed either by a post update or by another comment, and about 34\% of URCs remained unaddressed in our dataset. A strong majority (88.7\%) of the 417 addressed URCs are addressed in the next comment. Also, 139 comments (33.3\%) of these addressed URCs are addressed in the post body while 92 of them are addressed in the next comments as well.

\begin{table*}[h]
\caption{The main statistics about URCs in our dataset}
\vspace*{3mm}
\centering
\resizebox{14cm}{!}{%
\begin{tabular}{|l|c|c|}
\hline
\textbf{Statistical Questions} & \textbf{Count} & \textbf{Percent} \\
\hline
How many comments are URC? & 631 of 1221 & 51.7\% \\
\hline
How many of URCs are addressed (either in post or next comments)? & 417 of 631 & 66.1\% \\
\hline
How many of the addressed URCs are addressed in the next comments? & 370 of 417 & 88.7\% \\
\hline
How many of the addressed URCs are addressed in the post body? & 139 of 417 & 33.3\% \\
\hline
How many of the addressed URCs are addressed both in the next & 92 of 417 & 22.1\% \\
comments and the post body? &  & \\
\hline
\end{tabular}
\label{table-URC-statistics}
}
\end{table*}

\vspace{.2cm}
\noindent \textbf{RQ2: How fast are URCs addressed?}

Same as RQ1, we use the annotated dataset to answer this question. In our labeling, we didn't mention which post-update addresses the URC. But, we have labeled whether each URC is addressed by post-update or not. Thus, in case of being addressed in the post body, we assume the first post-update after that URC is the one that addresses it.   

Table~\ref{table-URC-addressed-within} shows the percentage of URCs that are addressed within 5 minutes, 1 hour, 1 day, 7 days, and a year. The table reports the portions based on both the 417 addressed URCs and all 631 URCs. Interestingly, 20\% of 631 URCs are addressed within 5 minutes, and about 55.3\% of them are addressed within a day. Among 417 addressed URC, 400 comments (95.9\%) were addressed within a year, meaning that 17 comments waited for more than a year to get addressed. On the other hand, among 47 unaddressed URC that were posted after 3/15/2019, one of them addressed within a year but after 3/15/2020. So, among 631 URCs, 417-17+1=401 comments (63.5\%) addressed within a year, and 230 comments (36.5\%) remained unaddressed within this period.

\begin{table}[h]
\caption{The percentage of addressed URCs within the specified times}
\vspace*{3mm}
\centering
\resizebox{10cm}{!}{%
\begin{tabular}{|c|c|c|}
\hline
\textbf{Addressed within} & \textbf{Of 417 addressed URCs} & \textbf{Of 631 URCs} \\
\hline
5 min & 30.2\% & 20.0\%  \\
1 hour & 58.8\% & 38.8\%  \\
1 day & 83.7\% & 55.3\%  \\
7 days & 87.5\% & 57.8\%  \\
1 year & 95.9\% & 63.4\%  \\
\hline
\end{tabular}
\label{table-URC-addressed-within}
}
\end{table}

\vspace{.2cm}
%\noindent \textbf{RQ3: Who addressed URCs in where?}
\noindent \textbf{RQ3: Which user role (questioner, answerer, other commenters) is more likely to address URCs? And in which part of the answer post do they choose to address URCs?}

%method
Same as previous RQs, we investigate our dataset to answer this question. Table~\ref{table-URC-who-addresses} shows the results. The rows show who (which user role) addressed the URCs, and the columns show where (which part of the post) URCs are addressed. There are 370 URCs that are addressed in the next comments, which a majority of them (296) are addressed by the answer owner (i.e., the person who posted the answer for the first time). The remaining are addressed by answer editor, i.e., the user who update the existing answer (3 URCs), by the questioner (15 URCs), and the other users (56 URCs).

\begin{table}[h]
\caption{Role of users who address the URCs and where they address the URCs}
\vspace*{3mm}
\centering
\resizebox{10cm}{!}{%
\begin{tabular}{|l|cccc|}
\hline
\textbf{URCs addressed} & in comment & in post & in either & in both \\
\hline
by answer owner  & 296 & 114 & 334 & 76  \\
by answer editor & 3   & 25  & 27   & 1   \\
by questioner    & 15   & 0  & 15   & 0  \\
by others        & 56  & 0  & 56  & 0  \\
\hline
by anyone        & 370 & 139 & 417 & 92 \\
\hline
\end{tabular}
\label{table-URC-who-addresses}
}
\end{table}

There are 139 post edits (to address URCs) which 114 of them are by the post owners and the remaining 25 post edits are by answer editors.

The answer owners addressed 334 URCs either in the post or in the next comments. It means that 334/417=80.1\% of addressed URCs are addressed by the answer owners. The table also shows that 76 URCs are addressed by the post owner both in the next comments and in the post body.

Note that the values in ``by anyone'' row for ``in either'' and ``in both'' columns are not the summation of that columns. For example, among 92 URCs that are addressed both in the next comments and in the post body, some of them are addressed by post owner in the post body (without writing a comment by them) while they also addressed by others in the comment.

\vspace{.2cm}
\noindent \textbf{RQ4: Can comment votes be used to distinguish URCs from non-URCs?}

Table~\ref{table-comment-scores-quantiles} shows the different quantiles of comment scores (votes) for each group. The comment score for more than 75\% of comments of each group is 0. Also, the quantiles for 80\% and 85\% are equal, indicating that the comment scores are not good means to detect URCs.

\begin{table}[htbp]
\caption{The quantiles of comment scores for each group of comments}
\vspace*{3mm}
\centering
\resizebox{8cm}{!}{%
\begin{tabular}{|l|c|c|c|c|c|c|}
\hline
\textbf{Category} & \textbf{50\%} & \textbf{75\%} & \textbf{80\%} & \textbf{85\%} & \textbf{90\%} & \textbf{95\%} \\
\hline
NO\_URC & 0 & 0 & 1 & 1 & 1 & 2 \\
\hline
URC & 0 & 0 & 1 & 1 & 2 & 4 \\
\hline
\end{tabular}
\label{table-comment-scores-quantiles}
}
\end{table}

\begin{tcolorbox}
\textbf{Summary:} About half of the comments are URC. While 55.3\% of URCs are addressed within 24 hours, 36.5\% of URCs remain unaddressed after a year. Also, the majority (80.1\%) of addressed URCs are addressed by the answer owner. The majority URCs' score is 0, which may not be visible to the community. 
\end{tcolorbox}
\vspace{-0.2cm}

\section{Automatically Detect Update Request Comments}\label{sec:URCDetection}
\subsection{URC Detection}\label{sec:urc-detection-subsec}

We know that SO notifies the owner of posts when someone writes a new comment on their post. This SO feature explains why most of the URCs (80.1\%) are addressed by the owner of the answering post, and why 55.3\% of the URCs are addressed within 24 hours, whereas 36.5\% remain unaddressed after a year. In addition, we observe that the majority URCs have a score of 0, which may cause them invisible to community members who are potentially interested in addressing them. Therefore, if there is a reliable predictive model to automatically detect URCs, Stack Overflow can apply it to improve post qualities by decreasing the number of unaddressed URCs. 
For example, when a new URC is posted on an answer and accurately identified by a URC detector, if the answer remains unchanged and no other comments appear after the identified URC, SO could predict this URC to be an unaddressed URC and push it to the community members who are interested in addressing the issue pointed out in the URCs, including the original answer. Thus, in this section we aim to investigate how URCs are predictable.

To predict URCs, we apply three classic ML models and two deep learning models (a CNN and a BERT based model). Each ML model utilizes either the features extracted from comments, or the TF-IDF of the comment text or both of them (i.e., the extracted features and the TF-IDF table that are horizontally concatenated). Each DL model utilizes either the text only or the text plus the features extracted from comments. We describe the details of the feature extraction process, TF-IDF extraction, and training classifiers as below.

\vspace{.2cm}

\noindent \textbf{Feature Extraction:}
Table~\ref{table-comment-features} describes the features we extracted for each comment by considering multiple dimensions, inspired by our manual comment annotation process. These dimensions are as follow:
\begin{itemize}
  \item \textbf{Comment features:} the features that are related to global aspects of a comment. It includes comment\_score and comment\_order.
  \item \textbf{Post features:} the features that are related to global aspects of a post. It includes post\_score and post\_comment\_count.
  \item \textbf{User features:} we believe the role of the user who posts a comment is an important clue to detect URCs. Most NO\_URCs are posted by the answerer to address previous URCs. We also consider user\_reputation that is provided by Stack Overflow. As users with high reputation have the permission of updating answers, they may update answers by themselves in case of new URCs, and write new comments when they want to address a URC.
  \item \textbf{Time features:} These features are the relative time between a comment and its previous/next comment and its previous/next post updates. The rational behind these features is that, in most cases URCs are addressed within a short period of time.
  \item \textbf{Text similarity features:} Each comment is more or less related to other comments and post updates. In most cases, they are related to immediately previous/next comment or the immediate next post update. We use two different text similarity measures, i.e., the Jaccard similarity and cosine similarity between the BERT vectors (obtained via running SBERT\footnote{https://www.sbert.net/} on each comment) to extract the similarities between a comment and its previous/next comments. We also use the Jaccard similarity to find the similarity between a comment and the next post change occurred after that comment. So, by extracting these similarities, we will find that how near events are related to the current comment.
  \item \textbf{Text semantic features:} We use TextBlob \citep{loria-2018}, a Python text processing library to extract the polarity and subjectivity of comments. We expect critic comments (that are kind of URCs) have a negative polarity.
  \item \textbf{Text extracted features:} There are some characters such as question mark, keywords such as ``exception'' or ``but'', URLs, and emotions (like ;)) that knowing their existence in a text provides good information about the content of a comment. We also considered the text length, because it somehow shows how much information is provided by that comment.
\end{itemize}

\noindent \textbf{TF-IDF Extraction:}
Term frequency–inverse document frequency (TF-IDF) is one of the most popular methods to vectorize documents of a corpus. A word in the TF-IDF of a document achieves a high value if it has a high frequency in that document but a low frequency in the whole collection of documents. To archive the best results, we remove the stop words and also the words with very low frequency, i.e., one or two.

\begin{table}[h]
\caption{The extracted features from each comment}
\vspace*{3mm}
\centering
\resizebox{13.5cm}{!}{%
\begin{tabular}{|ll|}
\hline
\textbf{Feature} & \textbf{Description} \\
\hline
\hline
\textbf{Comment Features:} & \\
comment\_score & The score of the comment which is zero or a positive number.\\ %Downvoting is not allowed for SO comments. \\
comment\_order & The order of the comment on that post. The first comment is 1, the second is 2, and so on. \\
\hline
\textbf{Post Features:} & \\
post\_score & The score of the post that can be a positive or negative number \\
post\_comment\_count & The number of comments on that post. \\
\hline
\textbf{User Features:} & \\
by\_asker & True: if the comment is written by the user who posted the question. False: otherwise. \\
by\_answerer & True: if the comment is written by the user who posted the answer. False: otherwise. \\
by\_not\_seen\_commenter & True: if the comment is written by a user that is neither the questioner nor the answerer, \\
 & and has not written any comment on this post before. False: otherwise. \\
by\_seen\_commenter & True: if the comment is written by a user that is neither the questioner nor the answerer, \\
& but has written at least one comment on this post before. False: otherwise. \\
user\_reputation & The reputation of the user who wrote the comment \\
\hline
\textbf{Time Features:} & \\
prev\_post\_edit\_time & log([The time of the comment] - [The time of the previous post edit] reported in minutes) \\
next\_post\_edit\_time & log([The time of the next post edit] - [The time of the comment] reported in minutes) \\
prev\_comment\_time & log([The time of this comment] - [The time of the previous comment] reported in minutes) \\
next\_comment\_time & log([The time of the next comment] - [The time of this comment] reported in minutes) \\
\hline
\textbf{Text Similarity Features:} & \\
prev\_comment\_jaccard\_sim & The Jaccard similarity between this comment and the previous comment \\
next\_comment\_jaccard\_sim & The Jaccard similarity between this comment and the next comment \\
prev\_comment\_bert\_sim & The Cosine similarity between the BERT vector of this comment and the previous comment \\
next\_comment\_bert\_sim & The Cosine similarity between the BERT vector of this comment and the next comment \\
comment\_post\_change\_sim & The Jaccard similarity between this comment and the post change after this comment \\
\hline
\textbf{Text Semantic Features:} & \\
polarity & The polarity of the comment text (between -1 and +1) \\
subjectivity & The subjectivity of the comment text (between 0 and 1) \\
\hline
\textbf{Text Extracted Features:} & \\
text\_len & The length (number of characters) of the comment text. \\
starts\_with\_@ & If the comment starts with @. For example: ``@John please explain your code.'' \\
contains\_question\_mark & If the comment contains a question (?) mark. \\
contains\_exclamation\_mark & If the comment contains an exclamation (!) mark. \\
contains\_but & If the comment contains the word “but”. \\
contains\_exception & If the comment contains the word “exception”. \\
contains\_url & If the comment contains a URL. \\
contains\_emotions & If the comment contains emotions like :) \\
talks\_to\_role & If this comment talks to a specific person by @. 0: doesn’t include @user, \\
& 1: talks to the questioner, 2: talks to the answerer, 3: talks to a commenter \\
\hline
\end{tabular}
}
\label{table-comment-features}
\end{table}

\vspace{.2cm}

\noindent \textbf{Training Classifiers:} We consider three basic classification models, i.e., random  forest, logistic regression, and naive bayes. These models are picked as they are most commonly used in prediction tasks in Software Engineering domain~\citep{Yang-2020}, and have shown acceptable performance in those tasks. We also consider two deep learning models: 

\noindent 1) a convolutional neural network (CNN) proposed by \citet{Qu-2019} that has the capability of incorporating the external features to the neural network (after the convolutional layers). To get the most from the CNN model, we modified its layers and tune its hyper parameters according to the 10\% validation set that is randomly selected from 1,221 java comments. In the original implementation of the CNN model, the authors used three convolutional layers. Also they directly concatenated the external features to the convoluted text. However, we found that if we decrease the number of convolutional layers from three to one, and also if we use a dense layer before concatenating the external features to the convoluted text we get better results. For the hyper parameters, we used the same optimizer (Adam) and same parameter values except for the learning rate that was 0.001 but we changed it to 0.0005.

\noindent 2) Multimodal-Toolkit, the deep learning model proposed by \cite{gu-2021} that gets text and tabular data and provides eight different architectures to combine the tabular data with a Transformer (such as BERT). We use bert-base-uncased as the Transformer in this toolkit. Our primitive investigations on that eight different architectures showed that we can achieve the highest performance by the architecture that embeds the text to a vector of size 768, then concatenate it with the output of a MLP which gets the tabular data (non-text data) and converts them to a vector of size 500. To be more specific, the MLP has one hidden layer with 10,000 nodes, and there are 500 nodes in its output layer. So, the concatenated vector has a size of 768+500=1268. For the hyper parameters, we use the default parameters and suggested values. To be more specific, the batch size is 16, the Adam learning rate is 5e-5, and the number of training epochs is 3.

\subsection{Experiments}\label{sec:urcwarn-experiments}
\noindent \textbf{Evaluation Metrics:} The automated identification of URCs can be treated as a standard binary classification task. Thus, we use standard evaluation metrics, i.e., precision, recall, F1-score, accuracy, and area under the curve (AUC) to evaluate our models. Precision $P$ measures the correctness of our models in predicting the type of a comment, i.e., whether the comment is a URC or not. A prediction is considered correct if the predicted type is the same as the actual type of the comment. Precision is calculated as the proportion of correctly predicted URCs. Recall $R$ measures the completeness of a model. A model is considered complete if all of update request comments are predicted to be URC. Recall is calculated as the proportion of actual URCs that were correctly predicted as such. F1-score is the harmonic mean of precision and recall, i.e., $(\frac{2*P*R}{P+R})$. Accuracy is the most intuitive performance measure and it is the ratio of correct predictions to the total predictions. The area under a receiver operating characteristic (ROC) curve, abbreviated as AUC, measures the overall performance of a binary classifier \citep{hanley-1982}. The AUC value is within the range [0.5–1.0], where 0.5 represents the performance of a random classifier and 1.0 corresponds to a perfect classifier.

\vspace{.1cm}
\noindent \textbf{Experiment Setup:}
We use the 1,221 labeled comments from the SO Java community to train the models. To test these models, we create two test datasets, one from SO JavaScript community, and another from SO Python community. There are three main reasons behind the cross-programming language evaluation setup. First, due to the time limit, similar to literature studies on SO posts, e.g., \cite{Soni-2019} and \cite{Tang-2021}, we do not consider all questions on Stack Overflow but those tagged with specific programming languages. We pick Java, Python, and JavaScript, as they are reported to be among the most popular programming languages~\footnote{https://survey.stackoverflow.co/2022/\#technology}. Second, it would be easier for us to label URCs in posts tagged with these programming languages, as we have enough domain knowledge to understand the background and content of the questions. Last but not least, we choose to perform cross-programming language prediction because URCs in different programming language communities might share different characteristics, and we would like to investigate if the predictive models trained from one programming language can perform stably in other programming languages. As all the extracted features shown in Table~\ref{table-comment-features} are language independent, all the applied ML and DL models in our experiments are language independent consequently. Therefore, we can train the models on the Java community comments, and test them on different datasets from different domains (e.g., JavaScript or Python), and expect similar performance on each domain. 

To create the test datasets, we tag 377 comments (posted on 100 randomly selected answers) from the SO JavaScript community and 289 comments (posted on 100 randomly selected answers) from the Python community. To randomly select 100 answering post from JavaScript or Python community, we follow the steps described in Algorithm 1, but using \verb!<javascript>! or \verb!<python>! tag.

As random forest is a stochastic algorithm, it provides different results in each run. So, we run this algorithm for 100 iterations and report the result with the median accuracy. In each iteration, we train the algorithm by the 1,221 comments from the Java community and test it on 377 comments from the JavaScript community and 289 comments from the Python community. For logistic regression and naive bayesian models, as they provide stable results, we only run them for once. As the CNN model provides stable results through different runs, we also run it for once. For Multimodal-Toolkit, we run it for 11 times and take the result with median accuracy.

Among the features mentioned in Table~\ref{table-comment-features}, six features may not be available in real scenarios: \verb!next_comment_jaccard_sim!, \verb!next_comment_bert_sim!, \verb!comment_post_change_jacc_sim!, \verb!next_post_edit_time!, \verb!comment_score!, and \verb!next_comment_time!. These features need some time to be available when a new comment is posted. So, we drop them before running the experiments. 

\vspace{.1cm}
\noindent \textbf{Baseline:} We also compare our models with the heuristic approach provided by \citet{Soni-2019} (details provided in Section~\ref{sec:relatedwork}). They use three heuristics that are based on regular expressions and the code parts that are common between comments and post updates.

The heuristic algorithm proposed by Soni and Nadi generates four labels, three of which are equivalent to our labels, but their UNKNOWN label is undefined in our labeling. Moreover, some comments are discarded by their algorithm. We decided to treat UNKNOWN and discarded comments in two different ways: One way is to treat all of them as NO\_URC. The second way is to drop all UNKNOWN or discarded comments and report the performance on the remaining comments. Among 377 JavaScript comments, 146 comments were discarded or labeled as UNKNOWN by their heuristic algorithm. For the Python community, among 289 comments, 101 comments were discarded or labeled as UNKNOWN by this heuristic.

\subsection{Results}\label{sec:results-subsec}
Table~\ref{table-performance-on-javascript} and Table~\ref{table-performance-on-python} show the performance of the three ML models (that each one is trained by three different inputs), the CNN model (trained with two different inputs), the Multimodal-Toolkit by \cite{gu-2021} that uses BERT (trained with two different inputs), and two baselines adjusted from Soni and Nadi's heuristic approach, on JavaScript and Python respectively. The results show that for both test data, the Multimodal-Toolkit trained by the extracted features + text achieves the best performance, i.e., about 90\% accuracy and 0.96 AUC. The random forest model trained by features+TF-IDF archives the second rank in the list for both test data.

The results in Table~\ref{table-performance-on-javascript} and Table~\ref{table-performance-on-python} also reveal that when we only use the text data (TF-IDF or pure text), none of the models achieves an accuracy higher than 74\%. However, when we only employ the extracted features, all of the three ML models achieve much higher performance comparing their TF-IDF based models. Also, for the deep learning based models (CNN and Multimodal-Toolkit), incorporating the extracted features resulted in a much higher performance comparing to its text based version. As expected, among the models which only uses the text data, either pure text or TF-IDF, the BERT based models provide the highest performance.

The two considered baselines did not perform well on our datasets. One potential reason is that regular expressions may not be as accurate as ML approaches. Moreover, we included additional important features such as the role of commenters. Finally, their definition of URC is not exactly the same as ours. The details are presented in section~\ref{sec:relatedwork-soni}.

\begin{table}[h]
\caption{The performance of different models with different input features on JavaScript comments} 
\vspace*{3mm}
\centering
\resizebox{14cm}{!}{%
\begin{tabular}{|c|c|c|c|c|c|}
\hline
\textbf{Classifier} & \textbf{Acc} & & & & \\
\textbf{(Input)} & \textbf{AUC} & \textbf{Category} & \textbf{P} & \textbf{R} & \textbf{F1}  \\
\hline
\hline
\textbf{RandomForest} & 88.3\% & NO\_URC & 0.886 & 0.867 & 0.876   \\
\cline{3-6} 
\textbf{(features)} & 0.946 & URC & 0.881 & 0.898 & 0.889 \\
\hline
\textbf{RandomForest} & 60.7\% & NO\_URC & 0.611 & 0.489 & 0.543   \\
\cline{3-6} 
\textbf{(TF-IDF)} & 0.655 & URC & 0.605 & 0.716 & 0.656 \\
\hline
\textbf{RandomForest} & 89.4\% & NO\_URC & 0.902 & 0.872 & 0.887   \\
\cline{3-6} 
\textbf{(features + TF-IDF)} & 0.949 & URC & 0.887 & 0.914 & 0.900 \\
\hline
\hline
\textbf{LogisticRegression} & 70.8\% & NO\_URC & 0.765 & 0.561 & 0.647   \\
\cline{3-6} 
\textbf{(features)} & 0.769 & URC & 0.678 & 0.843 & 0.751 \\
\hline
\textbf{LogisticRegression} & 65.0\% & NO\_URC & 0.638 & 0.617 & 0.627   \\
\cline{3-6} 
\textbf{(TF-IDF)} & 0.686 & URC & 0.660 & 0.680 & 0.670 \\
\hline
\textbf{LogisticRegression} & 84.1\% & NO\_URC & 0.895 & 0.756 & 0.819   \\
\cline{3-6} 
\textbf{(features + TF-IDF)} & 0.928 & URC & 0.804 & 0.919 & 0.858 \\
\hline
\hline
\textbf{GaussianNB} & 66.3\% & NO\_URC & 0.627 & 0.728 & 0.674   \\
\cline{3-6} 
\textbf{(features)} & 0.730 & URC & 0.708 & 0.604 & 0.652  \\
\hline
\textbf{GaussianNB} & 55.4\% & NO\_URC & 0.524 & 0.733 & 0.611   \\
\cline{3-6} 
\textbf{(TF-IDF)} & 0.600 & URC & 0.616 & 0.391 & 0.478  \\
\hline
\textbf{GaussianNB} & 66.3\% & NO\_URC & 0.627 & 0.728 & 0.674   \\
\cline{3-6} 
\textbf{(features + TF-IDF)} & 0.730 & URC & 0.708 & 0.604 & 0.652  \\
\hline
\hline
\textbf{CNN} & 66.8\% & NO\_URC & 0.655 & 0.644 & 0.650   \\
\cline{3-6} 
\textbf{(text)} & 0.706 & URC & 0.680 & 0.690 & 0.685  \\
\hline
\textbf{CNN} & 88.3\% & NO\_URC & 0.915 & 0.833 & 0.872   \\
\cline{3-6} 
\textbf{(text + features)} & 0.947 & URC & 0.859 & 0.929 & 0.893  \\
\hline
\hline
\textbf{Multimodal-Toolkit by \cite{gu-2021} using BERT} & 74.0\% & NO\_URC & 0.720 & 0.744 & 0.732   \\
\cline{3-6} 
\textbf{(text)} & 0.820 & URC & 0.759 & 0.736 & 0.747  \\
\hline
\textbf{Multimodal-Toolkit by \cite{gu-2021} using BERT} & \textbf{89.9}\% & NO\_URC & 0.908 & 0.878 & 0.893   \\
\cline{3-6} 
\textbf{(text + features)} & \textbf{0.953} & URC & 0.892 & 0.919 & 0.905  \\
\hline
\hline
\textbf{Heuristic by \citet{Soni-2019} (treat UNKNOWN and} & 49.3\% & NO\_URC & 0.480 & 0.744 & 0.584   \\
\cline{3-6} 
\textbf{discarded comments as NO\_URC)} &  & URC & 0.531 & 0.264 & 0.353  \\
\hline
\textbf{Heuristic by \citet{Soni-2019} (ignore UNKNOWN} & 50.2\% & NO\_URC & 0.481 & 0.582 & 0.527   \\
\cline{3-6} 
\textbf{and discarded comments)} &  & URC & 0.531 & 0.430 & 0.475  \\
\hline
\end{tabular}
\label{table-performance-on-javascript}
}
\end{table}

\begin{table}[h]
\caption{The performance of different models with different input features on Python comments}
\vspace*{3mm}
\centering
\resizebox{14cm}{!}{%
\begin{tabular}{|c|c|c|c|c|c|}
\hline
\textbf{Classifier} & \textbf{Acc} & & & & \\
\textbf{(Input)} & \textbf{AUC} & \textbf{Category} & \textbf{P} & \textbf{R} & \textbf{F1}  \\
\hline
\hline
\textbf{RandomForest} & 87.2\% & NO\_URC & 0.930 & 0.811 & 0.866   \\
\cline{3-6} 
\textbf{(features)} & 0.943 & URC & 0.825 & 0.936 & 0.877 \\
\hline
\textbf{RandomForest} & 63.0\% & NO\_URC & 0.672 & 0.541 & 0.599   \\
\cline{3-6} 
\textbf{(TF-IDF)} & 0.679 & URC & 0.600 & 0.723 & 0.656 \\
\hline
\textbf{RandomForest} & 87.9\% & NO\_URC & 0.931 & 0.824 & 0.875   \\
\cline{3-6} 
\textbf{(features + TF-IDF)} & 0.947 & URC & 0.835 & 0.936 & 0.883 \\
\hline
\hline
\textbf{LogisticRegression} & 72.3\% & NO\_URC & 0.788 & 0.628 & 0.699   \\
\cline{3-6} 
\textbf{(features)} & 0.786 & URC & 0.678 & 0.823 & 0.744 \\
\hline
\textbf{LogisticRegression} & 63.7\% & NO\_URC & 0.684 & 0.541 & 0.604   \\
\cline{3-6} 
\textbf{(TF-IDF)} & 0.694 & URC & 0.605 & 0.738 & 0.665 \\
\hline
\textbf{LogisticRegression} & 83.4\% & NO\_URC & 0.910 & 0.750 & 0.822  \\
\cline{3-6} 
\textbf{(features + TF-IDF)} & 0.933 & URC & 0.778 & 0.922 & 0.844 \\
\hline
\hline
\textbf{GaussianNB} & 64.4\% & NO\_URC & 0.615 & 0.811 & 0.700   \\
\cline{3-6} 
\textbf{(features)} & 0.741 & URC & 0.702 & 0.468 & 0.562  \\
\hline
\textbf{GaussianNB} & 59.5\% & NO\_URC & 0.578 & 0.777 & 0.663   \\
\cline{3-6} 
\textbf{(TF-IDF)} & 0.635 & URC & 0.633 & 0.404 & 0.494  \\
\hline
\textbf{GaussianNB} & 64.4\% & NO\_URC & 0.615 & 0.811 & 0.700   \\
\cline{3-6} 
\textbf{(features + TF-IDF)} & 0.741 & URC & 0.702 & 0.468 & 0.562  \\
\hline
\hline
\textbf{CNN} & 64.7\% & NO\_URC & 0.669 & 0.615 & 0.641   \\
\cline{3-6} 
\textbf{(text)} & 0.682 & URC & 0.627 & 0.681 & 0.653  \\
\hline
\textbf{CNN} & 87.5\% & NO\_URC & 0.889 & 0.865 & 0.877   \\
\cline{3-6} 
\textbf{(text + features)} & 0.953 & URC & 0.862 & 0.887 & 0.874  \\
\hline
\hline
\textbf{Multimodal-Toolkit by \cite{gu-2021} using BERT} & 74.0\% & NO\_URC & 0.735 & 0.770 & 0.752   \\
\cline{3-6} 
\textbf{(text)} & 0.826 & URC & 0.746 & 0.709 & 0.727  \\
\hline
\textbf{Multimodal-Toolkit by \cite{gu-2021} using BERT} & \textbf{90.7}\% & NO\_URC & 0.923 & 0.892 & 0.907   \\
\cline{3-6} 
\textbf{(text + features)} & \textbf{0.969} & URC & 0.890 & 0.922 & 0.906  \\
\hline
\hline
\textbf{Heuristic by \citet{Soni-2019} (treat UNKNOWN and} & 50.5\% & NO\_URC & 0.513 & 0.669 & 0.581   \\
\cline{3-6} 
\textbf{discarded comments as NO\_URC)} &  & URC & 0.490 & 0.333 & 0.397  \\
\hline
\textbf{Heuristic by \citet{Soni-2019} (ignore UNKNOWN} & 50.5\% & NO\_URC & 0.522 & 0.495 & 0.508   \\
\cline{3-6} 
\textbf{and discarded comments)} &  & URC & 0.490 & 0.516 & 0.503  \\
\hline
\end{tabular}
\label{table-performance-on-python}
}
\end{table}

\begin{figure}[htbp]
\centerline{\includegraphics[width=0.8\textwidth]{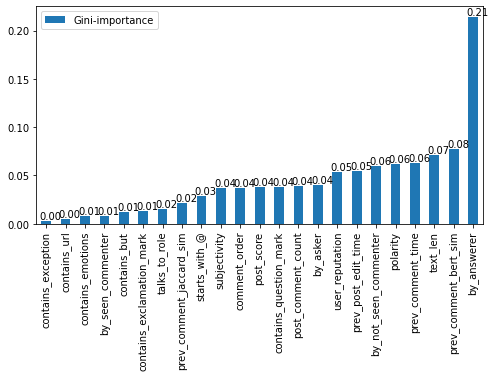}}
\vspace*{-3mm}
\caption{Feature importance by the random forest model}
\label{fig-feature-importance}
\end{figure}

Fig.~\ref{fig-feature-importance} shows the feature importance obtained by the feature based RF model with the median accuracy on JavaScript comments. As expected, \verb!by_answerer! has the highest weight because most of the addressing comments (that are NO\_URC) are written by the post answerer. The feature importance for other running iterations of RF is similar to this figure.

\begin{tcolorbox}
\textbf{Summary:} The update request comments are highly detectable. Utilizing the features we proposed to extract from comments, the supervised models achieved 89.9\% and 90.7\% accuracy on JavaScript and Python community respectively.
\end{tcolorbox}
\vspace{-0.2cm}

\section{Discussion: Can we automatically identify unaddressed update request comments?}\label{sec:discussion}
%\subsection{Can we automatically identify unaddressed update request comments?}

Identifying unaddressed URCs from existing SO comments might also help the SO community to improve the awareness of potential unresolved issues in existing answer posts. Thus in this discussion, we explore whether comment features proposed in Section~\ref{sec:URCDetection} can also be applied to identify unaddressed URCs in existing SO comments.

\noindent \textbf{Features and Model:} We reuse the features proposed in Table~\ref{table-comment-features} (including the six not available to new comments) and apply the Multimodal-Toolkit with BERT (as the best model to identify URCs) to detect all three classes: NO\_URC, URC\_ADDRESSED, and URC\_UNADDRESSED. We train the model on the 1,221 Java comments and test it on 377 and 289 comments from accepted answers to JavaScript and Python questions, respectively.

\noindent \textbf{Baseline:} We compare our results with the heuristic rule-based model provided by \citet{Soni-2019} (ref. Section~\ref{sec:relatedwork}). 

\noindent \textbf{Experiments and Evaluation:} Following the same method of performance evaluation which is described in Section~\ref{sec:urcwarn-experiments}, we run the Multimodal-Toolkit for 11 times and report median accuracy value. 

\noindent \textbf{Results:} Table~\ref{table-multimodal-perfomance-3class} shows the performance measures for each category for both JavaScript and Python communities. The accuracy on JavaScript and Python comments is 84.1\% and 84.8\% respectively. However, the model cannot provide a high f1-score for the URC\_UNADDRESSED class indicating the difficulty of detecting this type of comments.

\begin{table}[h]
\caption{The performance of Multimodal-Toolkit using BERT to detect 3-class comments in JavaScript and Python communities}
\vspace*{3mm}
\centering
\resizebox{10cm}{!}{%
\begin{tabular}{|c|c|c|c|c|c|}
\hline
\textbf{Community} & \textbf{Category} & \textbf{P} & \textbf{R} & \textbf{F1} & \textbf{Supp.} \\
\hline
\hline
 & NO\_URC & 0.869 & 0.922 & 0.895 & 180  \\
\cline{2-6} 
\textbf{JavaScript} & URC\_ADDRESSED & 0.868 & 0.797 & 0.831 & 148 \\
\cline{2-6} 
\textbf{(Acc: 84.1\%)} & URC\_UNADDRESSED & 0.660 & 0.673 & 0.667 & 49 \\
\hline
\hline
 & NO\_URC & 0.905 & 0.905 & 0.905 & 148  \\
\cline{2-6} 
\textbf{Python} & URC\_ADDRESSED & 0.827 & 0.910 & 0.867 & 100 \\
\cline{2-6} 
\textbf{(Acc: 84.8\%)} & URC\_UNADDRESSED & 0.645 & 0.488 & 0.556 & 41 \\
\hline
\end{tabular}
\label{table-multimodal-perfomance-3class}
}
\end{table}

Table~\ref{table-soni-perfomance} reports the performance of the heuristic model on JavaScript and Python comments when we treat the discarded and UNKNOWN comments as NO\_URC. It provides 39.8\% and 40.1\% accuracy on JavaScript and Python comments respectively that is far lower than the performance of our proposed model. As treating the discarded and UNKNOWN comments by the second way (i.e., ignoring them) led to even worse performance (i.e., 34.6\% and 34.6\%), we don't present the detailed performance obtained by this treatment. The above results show that our model outperforms the baseline with a large margin in terms of accuracy and f1-score.

\begin{table}[h]
\caption{ The performance of the heuristic model \citep{Soni-2019} on JavaScript and Python comments}
\vspace*{3mm}
\centering
\resizebox{10cm}{!}{%
\begin{tabular}{|c|c|c|c|c|c|}
\hline
 & \textbf{Category} & \textbf{P} & \textbf{R} & \textbf{F1} & \textbf{Supp.} \\
\hline
\hline
\textbf{} & NO\_URC & 0.480 & 0.744 & 0.584 & 180  \\
\cline{2-6} 
\textbf{JavaScript} & URC\_ADDRESSED & 0.250 & 0.047 & 0.080 & 148 \\
\cline{2-6} 
\textbf{(Acc: 39.8\%)} & URC\_UNADDRESSED & 0.129 & 0.184 & 0.151 & 49 \\
\hline
\hline
\textbf{} & NO\_URC & 0.513 & 0.669 & 0.581 & 148  \\
\cline{2-6} 
\textbf{Python} & URC\_ADDRESSED & 0.412 & 0.070 & 0.120 & 100 \\
\cline{2-6} 
\textbf{(Acc: 40.1\%)} & URC\_UNADDRESSED & 0.127 & 0.244 & 0.167 & 41 \\
\hline
\end{tabular}
\label{table-soni-perfomance}
}
\end{table}

Thus, we conclude that, \textbf{unlike URCs, URC\_UNADDRESSED are difficult to identify based on comment features.} In the future, more advanced models are needed to better capture the answer edit and comment post history.

\section{Threats to Validity}\label{sec:threats}
In this section, we start with threats to external validity. Our empirical study is based on a dataset that contains 1,221 labeled comments extracted from a set of statistically sampled answer posts from Java questions. Although Java is one of the biggest communities in SO, and we can expect similar results for other popular languages, our findings may not be the same for questions on other topics.  

Similarly, we only evaluate the performance of our models on sampled comments in accepted answer posts to questions relevant to Python, and JavaScript. Thus, the reported performance may not generalize to other communities. However, as the reported results across different categories are close in terms of accuracy, AUC and F1-score, we believe that our model should work for other communities. 

Another threat to external validity is the limited size of empirical study and evaluation datasets due to the time-consuming annotation process. In the empirical study, we manually annotated a statistically significant sample of Stack Overflow answer posts and their comments. Specifically, We randomly sampled 384 answer posts, achieving a 5\% margin error with a confidence level of 95\%. To evaluate the proposed automated approaches, we created two test datasets by manually annotating 289 comments that were posted on 100 Python posts and 377 comments that were posted on 100 JavaScript posts. The performance of our models on the two datasets are close to each other. But still, the above results may not be applied to all Stack Overflow posts and comments.

As for threats to internal validity, the empirical study and the train/test of automated URC detection models rely on manually labeled comments and might be biased or error-prone. To avoid such a problem, two authors independently checked 100 comments and reported a high agreement ratio. Therefore we believe our coding guide is clear and could support future replication on other data.

\section{Related Work}\label{sec:relatedwork}

\subsection{Studies on Stack Overflow Comments}\label{sec:relatedwork-soni}

The most relevant work to our study is by \citet{Soni-2019} that analyzed how comments affect answer updates on Stack Overflow, although they did not perform any empirical study. They employed a heuristic rule-based approach to classify comments into four categories. Since these categories are also determine whether a comment is asking for update, we clarify the relationship between their comment categories and our categories below to reduce ambiguities.

\begin{enumerate}
\item \textbf{WARRANT UPDATE}: A comment that warranted an update but an edit was not made to the answer. The equivalent of this class in our classification is URC\_UNADDRESSED.

\item \textbf{UPDATE}: A comment that warranted an update and an edit were made to the answer. This class is equivalent to our URC\_ADDRESSED.

\item \textbf{NO UPDATE}: A comment that did not warrant an update. It is equivalent to our NO\_URC class.

\item \textbf{UNKNOWN}: All other comments that are just text, URLs, or discussion (e.g., ``Thank you so much for this answer.''). 
\end{enumerate}

Soni and Nadi claimed that ``\textit{only a few ($\sim$4-5\%) comments resulted in answer updates}'' and ``\textit{More than a quarter ($\sim$26-29\%) of the comments we studied across the five tags} [java, python, javascript, android, php] \textit{require the answer to be updated, but are ignored by answer posters}''. From these statements, one can conclude that $\sim$84-88\%\footnote{26/(26+5)=83.9\% and 29/(29+4)=87.9\%} of update request comments are unaddressed. Different from them, by manually analyzing 1,221 comments, we found that only 36.5\% of URCs remain unaddressed after a year. Such inconsistent results may arise due to several reasons. First, their study scope is different from ours. They focused on the updates in the code block of an answer post and ignored the update in text. But in reality, answer owners sometimes only need to update the text to address URCs rather than touching code. We argue that text updates are also important because both text and code are important for a high-quality post \citep{Calefato-2015}, and developers rely on both when utilizing SO \citep{Wu-2019,Chatterjee-2020}. Secondly, Soni and Nadi ignore the answers in the following comments. However, we found that 58.6\% of our annotated URCs are addressed in the following comments. Thirdly, their findings rely on their rule-based approach to identify update-introduce comments, and their reported accuracy is 85\% on labeled comments from 30 answer posts. Instead, we manually examined the studied dataset, which is more accurate. 

Though their categorization is different from ours, we can still compare with their heuristic approach in identifying URCs and unaddressed URCs. Our experiment results (ref. Table~\ref{table-performance-on-javascript}, \ref{table-performance-on-python}, \ref{table-multimodal-perfomance-3class}, and ~\ref{table-soni-perfomance}) show that our proposed machine learning-based approach outperforms their approach with a large margin. Moreover, while the algorithm by \citet{Soni-2019} discards many comments due to either not including code updates or being labeled as UNKNOWN, our model labels all comments and doesn't discard any comment.

Another closed study on SO comments is by \citet{Zhang-2021}. They found that 4.4 million comments (possibly including informative comments) are hidden by default from developers. To help identify informative comments, they propose a taxonomy to group comments into seven types: Praise (praise an answer), Advantage (discuss the advantage of an answer), Improvement (make improvement to an answer), Weakness (point out the weakness of an answer), Inquiry (make inquiry based on an answer), Addition (provide additional information to an answer), and Irrelevant (discuss irrelevant topics to an answer). Their categorization is different from ours because our categorization focuses on whether a comment asks for an update on the answer post.

\subsection{Other Studies on Stack Overflow}
Studies on Stack Overflow can be categorized into two types, i.e., mining questions and answers on Stack Overflow to extract the challenges faced by developers~\citep{Treude-2011,Barua-2014, Rosen-2016,Ahmed-2018,Tahir-2018, Bagherzadeh-2019,Openja-2020,Tan-2020,Wen-2021} and investigating the mechanisms used by Stack Overflow and proposing new feature/model to improve user experience on Stack Overflow~\citep{Xia-2013,Saha-2013,Nasehi-2012,Asaduzzaman-2013,Ponzanelli-2014,Beyer-2015,Zhang-2015,Ahasanuzzaman-2016,Srba-2016,Yang-2016, Mizobuchi-2017, Chen-2018, Wang-2018,Wang-2018-2,Zhang-2019,Chatterjee-2020}.

\citet{Treude-2011} manually categorized the types of questions on Stack Overflow, and observed that Stack Overflow could be useful for code review and learning the concepts of programming. \citet{Barua-2014} first applied LDA, a popular statistical topic model, to discover topics from the contents on Stack Overflow and track the changes of the topics over time. Following their methodology, researchers have analyzed contents on Stack Overflow related to fine-grained domains. For instance, \citet{Rosen-2016} analyzed 13,232,821 posts to examine what mobile developers ask about. They discovered hot topics and determined what popular mobile-related issues are the most difficult. \citet{Ahmed-2018} applied a similar methodology to analyze what do concurrency developers ask on Stack Overflow. \citet{Tahir-2018} found that developers widely use Stack Overflow to ask for general assessments of code smells or anti-patterns instead of asking for particular refactoring solutions. More recent, Stack Overflow content related to big data analysis~\citep{Bagherzadeh-2019}, release engineering~\citep{Openja-2020}, bug severity~\citep{Tan-2020}, and serverless computing~\citep{Wen-2021} are analyzed to help relevant stakeholders better understand the trends, challenges, and potential future development/research directions.

Many prior studies are investigating the quality of the crowd-sourced knowledge presented on Stack Overflow. \citet{Asaduzzaman-2013} analyzed unanswered questions on Stack Overflow and found that the quality of questions is strongly related to whether a question receives an answer. \citet{Srba-2016} observed that an increasing amount of content with relatively lower quality is hurting the Stack Overflow community. \citet{Nasehi-2012} examined code examples on Stack Overflow and identified characteristics of high-quality code examples. \citet{Yang-2016} focused on the quality of code snippets on Stack Overflow. They examined the usability of code snippets by compiling or running them. \citet{Zhang-2019} analyzed the obsoleteness of answers on Stack Overflow. They found that more than half of the obsolete answers were probably already obsolete when they were first posted. Moreover, when an obsolete answer is observed, only a small proportion (20.5\%) of such answers are ever updated. Thus they suggest that Stack Overflow should develop mechanisms to encourage the whole community to maintain answers. \citet{Chatterjee-2020} conducted an exploratory study of novice software engineers’ focus in stack overflow posts. They found that Novice programmers focus on 15–21\% text and 27\% code in a Stack Overflow post.

Prior studies also examined Stack Overflow's mechanisms to understand its operation better and proposed tools to improve the efficiency of the knowledge-sharing process. For instance, to enhance the quality of knowledge on Stack Overflow, \citet{Ponzanelli-2014} proposed an automated approach to identify the quality of posts and filter low-quality content. \citet{Wang-2018-2} studied how Stack Overflow users revise answers and what is the impact of those revisions. They found that although the current badge system on Stack Overflow is designed to ensure the quantity of revisions, such a badge system fails to consider the quality of revisions and should be improved in the future. \citet{Chen-2018} proposed a convolutional neural network (CNN) based approach to predict the need for post revisions to improve the overall quality of Stack Overflow posts. Several approaches are raised to automatically predict tags on Stack Overflow questions~\citep{Xia-2013,Saha-2013,Beyer-2015,Wang-2018,Chen-2019}, and identify duplicate posts~\citep{Zhang-2015,Ahasanuzzaman-2016,Mizobuchi-2017}.

\section{Conclusion and Future Work}\label{sec:conclusion}
Comments on Stack Overflow answer posts act as a potential way to improve the quality of the answers, which is one main concern of Stack Overflow community. In this paper, we conduct a study on URCs (update request comments)---comments in answer posts that explicitly or implicitly ask for an answer update due to reasons such as warning issues in the answer. Specifically, we investigate what happens when a user posts a URC and how/when/by whom it gets addressed. For this purpose, we manually examine a sample set of 1,221 comments on answer posts of questions tagged with ``java''. We find that 51.7\% of the analyzed comments are URCs. Most addressed URCs (80.1\%) are addressed by the answer owners, and more interestingly, most URCs (55.3\%) are addressed within 24 hours. Nevertheless, 36.5\% of URCs remain unaddressed after a year. 

Upon checking the votes received by URCs, we find that the majority URCs have a score of 0, which may cause them to be invisible to community members who are potentially interested in addressing them. Thus, as the first step towards improving the awareness of URCs, we explore the feasibility of building a tool that can automatically identify URCs as they post. Such a tool can also be leveraged to mine URCs for research purposes. Specifically, we proposed a set of comment features for URC detection and trained several supervised models, including random forest and BERT from 1,221 annotated Java comments. We evaluated the performance of our models on Python and JavaScript comments. Experiments results show that our automated URC detector can identify URCs with around 90\% accuracy. 

In the future, we would like to increase the number of annotated comments for train and evaluation and investigate if specific kinds of URCs are more likely to be addressed. We also plan to analyze URCs that non-answer owner addresses to explore what types of SO users are more likely to help the community address URCs.

\section*{Acknowledgement}
We acknowledge the support of the Natural Sciences and Engineering Research Council of Canada (NSERC), [funding reference number: RGPIN-2019-05071].

\bibliographystyle{elsarticle-harv} 
\bibliography{main}





% \end{thebibliography}

\clearpage
\end{document}